# Angstrom-wide conductive channels in black phosphorus by Cu intercalation

**Authors:** Suk Woo Lee[1,2], Lu Qiu[1,2], Jong Chan Yoon[1,2], Yohan Kim[2], Da Li[1], Inseon Oh[2], Gil-Ho Lee[3], Jung-Woo Yoo[2], Hyung-Joon Shin[1,2], Feng Ding[1,2]* and Zonghoon Lee[1,2]*

**Affiliations:**

[1]Center for Multidimensional Carbon Materials, Institute for Basic Science (IBS), Ulsan 44919, Republic of Korea

[2]Department of Materials Science and Engineering, Ulsan National Institute of Science and Technology (UNIST), Ulsan 44919, Republic of Korea

[3]Department of Physics, Pohang University of Science and Technology (POSTECH), Pohang 37673, Republic of Korea

*Correspondence to: zhlee@unist.ac.kr, f.ding@unist.ac.kr

## Abstract

Two-dimensional (2D) materials have received much attention in view of their superior properties and their exotic behavior both of which result from their atomic scale thickness[1,2]. To further improve and modulate these properties, various methods including substitutional doping, functionalization, and defect engineering have been explored[3-5]. Among these, intercalation has numerous advantages in that new optical, magnetic, electronic or even superconducting properties can be induced while minimizing structural damage to the 2D matrix by defect formation[6-9]. Moreover, controlled intercalation of 2D materials where the desired intercalant can be spatially separated in a predesigned orderly manner can be potentially used to tailor material properties at atomic scale. Even so, there are few reports describing spatially controlled intercalation in a 2D matrix due to the difficulty in controlling the diffusion of intercalants[6]. Here, we report on the formation of angstrom-wide conductive channels (~4.3 Å) in black phosphorus (BP) by Cu intercalation. Due to inherent anisotropy in the buckled atomic structure of BP, Cu atoms anisotropically intercalate in BP along the zigzag direction occupying angstrom-wide spaces with Cu intercalated regions sandwiched between pristine BP regions. Using *in situ* experimental techniques combined with theoretical calculations, we reveal the intercalation process in detail with respect to atomic structure, microstructural modifications such as strain and deformation, and also elucidate the mechanism of intercalation. Conductive atomic force microscopy results point to localized differences in electronic properties with Cu intercalated regions having enhanced conductivity and showing semimetallic behavior. Our findings throw light on the fundamental relationship between microstructure changes and properties in intercalated 2D materials. Our study thus provides a new way to tailor the properties of anisotropic 2D materials at angstrom scale.

## Main

**Atomic structure of angstrom-wide Cu intercalated black phosphorus**

Two dimensional (2D) black phosphorus (BP) has an anisotropically puckered structure where the structural organization is distinctly different along the zigzag and armchair directions[10-12]. We anticipated this structural anisotropy of BP to profoundly influence intercalation in BP. To intercalate Cu into BP, Cu was first deposited on BP, followed by a heat treatment to diffuse the deposited Cu into the BP matrix (Fig. 1a). The Cu intercalated BP sample was imaged by high-angle annular dark-field scanning transmission electron microscopy (HAADF-STEM). Fig. 1b and c show, respectively, the plan-view and cross-sectional view of Cu intercalated BP where we identify Cu intercalated regions by the vertical white lines in the two figures; the higher intensity of the HAADF-STEM image in these regions is due to the higher atomic number of Cu. The white lines indicate that Cu atoms are anisotropically intercalated along the zigzag direction of BP.

Apart from atomic resolution HAADF-STEM analysis, energy dispersive X-ray spectroscopy (EDS) and density functional theory (DFT) calculation were used to further elucidate the atomic structure of Cu intercalated BP (Fig. 2). First, the Cu intercalated BP structure was analyzed along the four different zone axes of BP (as shown in the reconstructed image in Fig. 2a), after which the corresponding atomic resolution EDS maps were obtained (Fig. 2h and i). Based on experimental results, an atomic model of the Cu intercalated structure was then constructed and DFT calculation was used to relax structure. The simulated HAADF-STEM images of the final atomic model agree with experimental images. A slight difference between experimental and calculated results is the undulating morphology of the calculated image as compared to the straight lines observed in experimental images (Fig. 2c and d), although undulations could intermittently be observed in some experimental images too. We attribute this difference to the following: in our DFT calculation, all boundaries were allowed to move freely to obtain a fully relaxed structure. However, under real conditions, most intercalated regions could not be fully relaxed and hence show a straight rigid morphology in HAADF images unlike the calculated structures. In the unrelaxed structure, we expect induced local strain in the vicinity of intercalated regions. We will discuss this strain effect further in the following section.

In the Cu intercalated structure, we see periodically missing Cu atoms along the [101] zone axis of BP (Fig. 2c); every fifth Cu atom is missing and the column of missing Cu has an angle of ~73 ° with respect to the BP layer (Fig. 2g). Besides every fifth missing Cu, other missing atom periodicities were also occasionally observed. The coexistence of different periodicities indicates that Cu intercalated BP cannot be considered as a compound with a fixed atomic ratio of the two constituent elements. To explain above observation, we calculated Cu binding energies for several different periodicities of missing Cu and the result shows that the structure with every fifth Cu atom missing is energetically more favorable than others.

**Microstructural effects of Cu intercalation in black phosphorus: strain, deformation and atomic configurations in single crystalline black phosphorus**

In the previous section on the structure of Cu intercalated BP, the presence of strain was inferred by comparing the undulating and straight morphologies of calculated and experimental images of Cu intercalated structures (Fig. 2c, d). The Cu intercalated structure has a larger atomic distance than pristine BP, this increase being ~6.65 % in the zigzag direction and ~4.06% for interlayer distance (Fig. 3a). Since the BP matrix is restricted, this expansion cannot be compensated by a full relaxation of the structure, resulting in a straight rigid morphology in contrast to the undulating morphology derived from DFT calculations based on a fully relaxed structure (Fig. 2c, d). The strain induced by the unrelaxed Cu intercalated BP structure could be readily identified in and around Cu intercalated regions (Fig. 3a) due to which layers of BP in these regions appear to be fluctuating and the atomic structure is not clearly resolved. Interestingly, it turns out that the strain can be relaxed to some extent by forming kinks in the Cu intercalated structure. Thus, fluctuation and less resolved atomic structure of BP are predominantly observed in areas preceding the formation of a kink (the area just above the kink point in Fig. 3a) rather than after the kink (below the kink point in Fig. 3a). Consequently, as the density of Cu intercalation is increased, kinks are more frequently observed since the strain readily accumulates in high-density regions of the Cu intercalated structure followed by relaxation. This observation demonstrates that Cu intercalation induces strain in BP and this strain can be relaxed by the formation of kinks in the Cu intercalated regions.

In addition to structural strain, we found that Cu intercalation of BP introduced a layer mismatch as shown in the HAADF-STEM image of the [100] zone axis in Fig. 2f. This mismatch indicates that Cu intercalation deforms the structure by ~0.21 nm in the direction perpendicular to the BP layers. The DFT-calculated atomic model (Fig. 2f) more clearly illustrates the deformation of BP following Cu intercalation. The overall change in morphology of a single crystalline BP flake with a high density of Cu intercalation is illustrated in the low mag image in Fig. 3b. Structural deformation causes a buckling of the flake as a result of which some regions are tilted upwards and some slope downwards. Atomic resolution HAADF-STEM images of these regions show the presence of two inequivalent atomic configurations of Cu intercalated structures having mirror symmetry with respect to each other (bottom of Fig. 3b). Considering the layer mismatch of a single Cu intercalated structure (~0.21 nm), we can write the following equation to describe the deformation on a macroscopic scale:

$$D \cong 0.21 \cdot |n_{mirror\ A} - n_{mirror\ B}|\ nm \qquad (1)$$

Where $D$ is the height difference caused by deformation resulting from Cu intercalation at the two end surfaces of the region of interest, and $n_{mirror\ A\ (B)}$ is the number of Cu intercalated structures; $mirror\ A$ and $B$ indicate the two inequivalent atomic configurations which are related to each other by mirror symmetry (bottom of Fig. 3b). These results illustrate how microscopic deformation induced by Cu intercalation influences the macroscopic morphology of the BP crystal.

The mirror symmetry in Cu intercalated structure described in Fig. 3b implies that different atomic configurations are possible for the Cu intercalated structure. Further examination of these structures showed that Cu intercalated BP structure can have four different atomic configurations in single crystalline BP, all of which can be superimposed on each other through rotation.

**Rate-limiting Cu intercalation mechanism: Top-down intercalation**

To elucidate the mechanism of Cu intercalation in BP, we investigated truncated Cu intercalated structures where intercalation was interrupted before completion to the bottom of the BP flake (Fig. 4a and b). Two types of truncated structures were seen: in the first, two Cu atoms are fully intercalated at the end of the truncated structure (Fig. 4a) and in the second, there is one additional intercalated Cu atom in the partially intercalated next layer as compared to the structure in Fig. 4a (Fig. 4b). By repeating these two steps, Cu atoms can be intercalated from top to bottom of the BP crystal. In addition, as shown in Fig. 4c, Cu intercalation that does not contact with the edges of the BP flake was also frequently. This is a further indication that top-down intercalation is the dominant mechanism for Cu intercalation in BP.

We calculated the energy barriers of Cu diffusion along zigzag, armchair directions of a BP layer, and the crossing a BP layer by DFT calculations. The energy barriers of Cu diffusion along the armchair and zigzag directions (0.18 eV and 0.38 eV, respectively) are much lower than that crossing a BP layer (1.78 eV), although the latter energy barrier is comparable to that observed for vacancy-assisted top-down intercalation of $MoS_2$[13]. When strain was taken into account, the calculated diffusion barrier of crossing a BP layer was reduced (1.48 eV) (Fig. 4d-f and), indicating that top-down intercalation of Cu is more favorable nearby a intercalated region rather than that in a pristine BP region. We carried out *in situ* HAADF-STEM to experimentally investigate the mechanism of Cu intercalation (Fig. 4g, h). The measured rates of Cu intercalation at 180, 220, and 250 ºC were ~1.19, 18.91, and 78.10 nm/s, respectively, from which the activation energy of Cu intercalation ($E_a$) can be estimated by fitting the following formula

$$k = k_0 \cdot e^{\frac{-E_a}{k_B T}} \tag{3}$$

where $k$ is the rate of intercalation and $k_0$ is a constant, $k_B$ is the Boltzmann constant and $T$ is the temperature (Fig. 4h). The estimated activation energy (1.24 eV) agrees well with the calculated energy barrier of Cu diffusion crossing the BP layers (1.48 eV) and thus validated the proposed top-down mechanism of intercalation. This result implies that the overall intercalation process is limited by the diffusion of Cu atoms crossing BP layer and we therefore conclude that top-down Cu intercalation is the rate-limiting process.

**Angstrom-wide conductive channels by Cu intercalation in black phosphorus**

According to previous reports, doping with a transition metal such as Cu in BP generally causes a Fermi level shift without significantly changing the electronic band structure[14,15]. These results were obtained however, by assuming a random distribution of the doped metal atom in BP, resulting in a lower estimate of the metal atom concentration in the BP matrix. Our research has shown that Cu atoms intercalate periodically into BP. Thus, we expect the electronic band structure to be significantly changed in the Cu intercalated regions. Fig. 5b-c show the calculated density of states (DOS) and electronic band structure of Cu intercalated BP. Semimetallic behavior is clearly evidenced in Cu intercalated BP in contrast to the semiconducting nature of pristine BP.

To experimentally investigate local variations in electronic properties in Cu intercalated BP, we performed conductive atomic force microscopy (C-AFM) (Fig. 5d and e). The Cu intercalated regions were easily identified in morphology images since Cu intercalation leads to a deformation-induced height difference. We have already noted from our microstructure studies that Cu intercalation induced a height difference (Fig. 2f and Fig. 3b). In the current mapping image, higher conductivities were measured at Cu intercalated regions than in the pristine areas of BP (Fig. 5d). The higher conductivity was repeatedly observed in all Cu intercalated regions. Comparing the *I-V* curves at the Cu intercalated and pristine regions also confirmed the higher conductivity at the Cu intercalated regions (Fig. 5e). The locally increased conductivity agrees with the calculated electronic properties where semimetallic property appears at the Cu intercalated region while pristine BP remains semiconducting. These results demonstrate that angstrom-wide conductive channels can be formed in anisotropic 2D materials by a simple intercalation method.

In this study, we reveal that Cu atoms are anisotropically and periodically intercalated in BP to form angstrom-wide conductive channels. The atomic structure and the resultant microstructural effects, including strain and deformation, are described at atomic scale. Our *in situ* experiments and computational calculations show that Cu intercalation follows a top-down intercalation mechanism which is the rate-limiting process. Finally, calculated electronic properties and C-AFM directly point to local variations in electronic properties in Cu intercalated regions which have higher conductivity than unintercalated BP. Thus, we show that intercalation can be used to modulate various properties in BP and intercalation could be a versatile new way to tailor the properties of anisotropic 2D materials at angstrom scale.

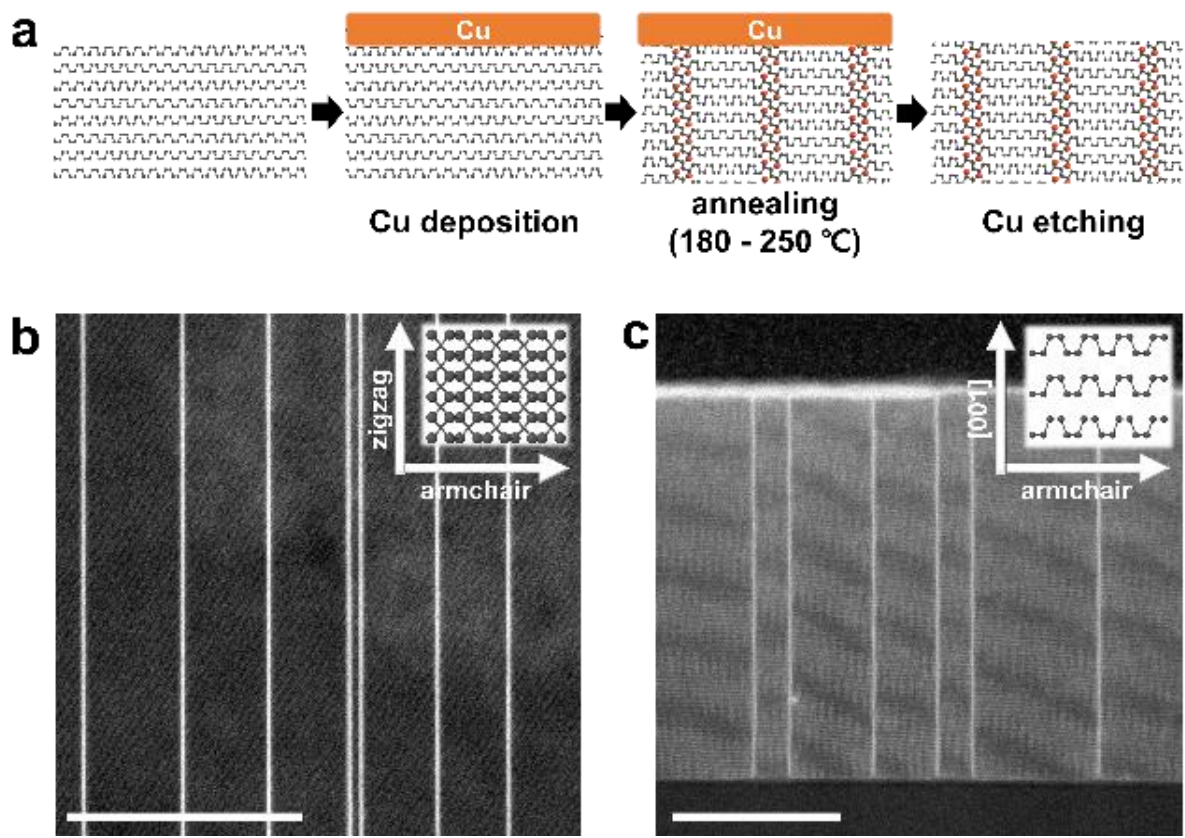

**Fig. 1 | Anisotropic Cu intercalation in BP. a**, Schematic diagram of the method for Cu intercalation in BP. **b**, **c**, High-angle annular dark-field scanning transmission electron microscopy (HAADF-STEM) images in plan-view (**b**) and cross-sectional view (**c**) of BP after Cu intercalation. Scale bars are 50 nm.

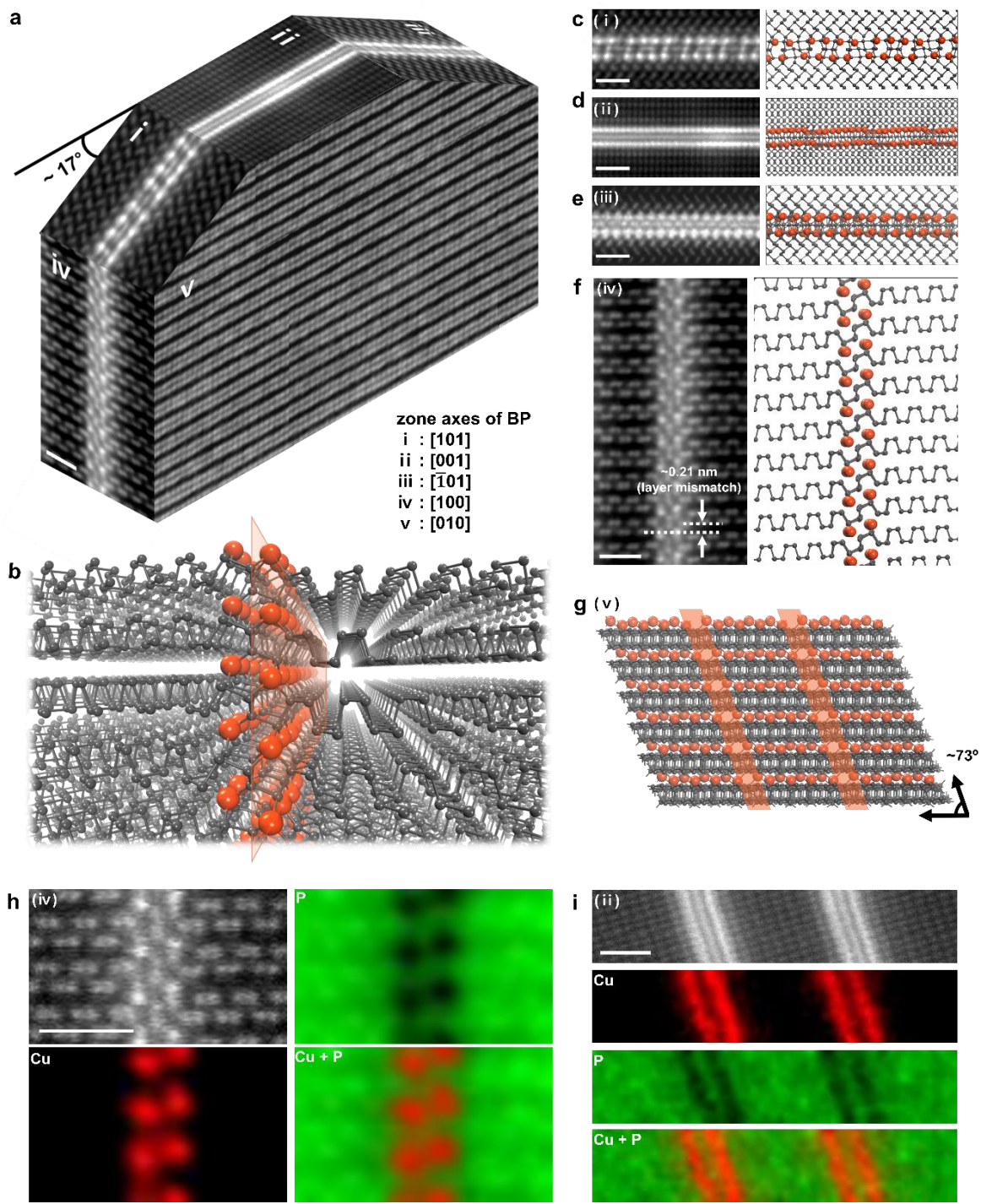

**Fig. 2 | Atomic structure of Cu intercalated BP. a**, Three-dimensional image constructed from atomic resolution HAADF-STEM images of Cu intercalated BP at five different zone axes of BP. **b**, Perspective atomic model of Cu intercalated BP. **c-f**, HAADF-STEM images (left) with the corresponding atomic models (right) calculated from density functional theory (DFT) at the four zone axes of BP ((**c**) [101], (**d**) [001], (**e**) [1̄01], (**f**) [100]). Scale bars are 1 nm. **g**, Atomic model of Cu intercalated BP at [010] zone axis of BP. **h,i**, Atomic resolution energy dispersive X-ray spectroscopy (EDS) mappings at [100] (**h**) and [001] (**i**) zone axes of BP. Scale bars are 1 nm.

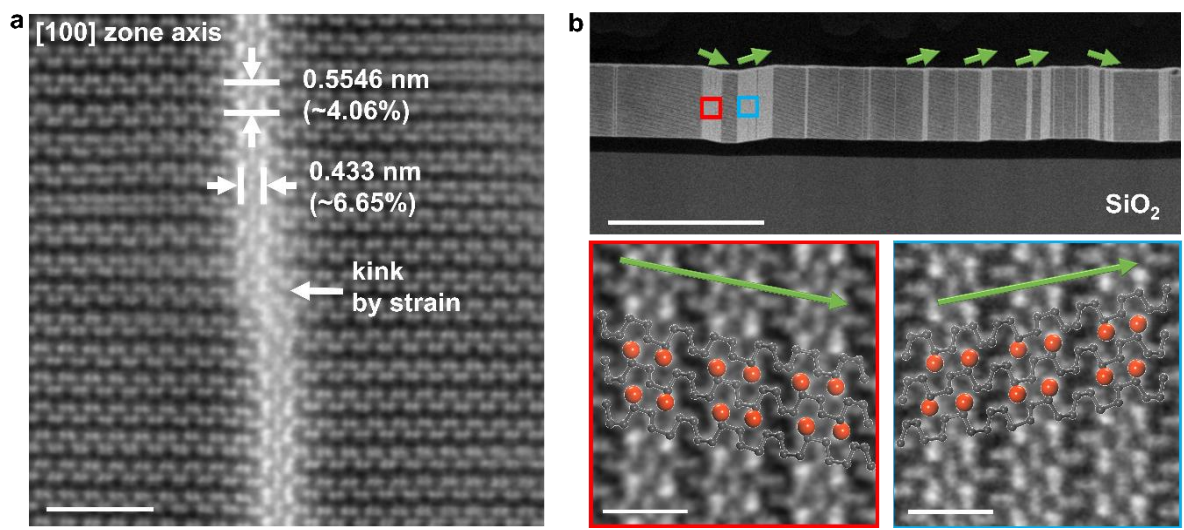

**Fig. 3 | Microstructural effects of Cu intercalation in BP. a**, Atomic resolution HAADF-STEM image of strain-induced kink formation in a Cu intercalated structure ([100] zone axis). Scale bar is 2 nm. **b**, Low magnification (top) and atomic resolution (bottom) HAADF-STEM images of densely Cu intercalated BP in cross-sectional view ([100] zone axis). Scale bars are 200 nm (top) and 1 nm (bottom).

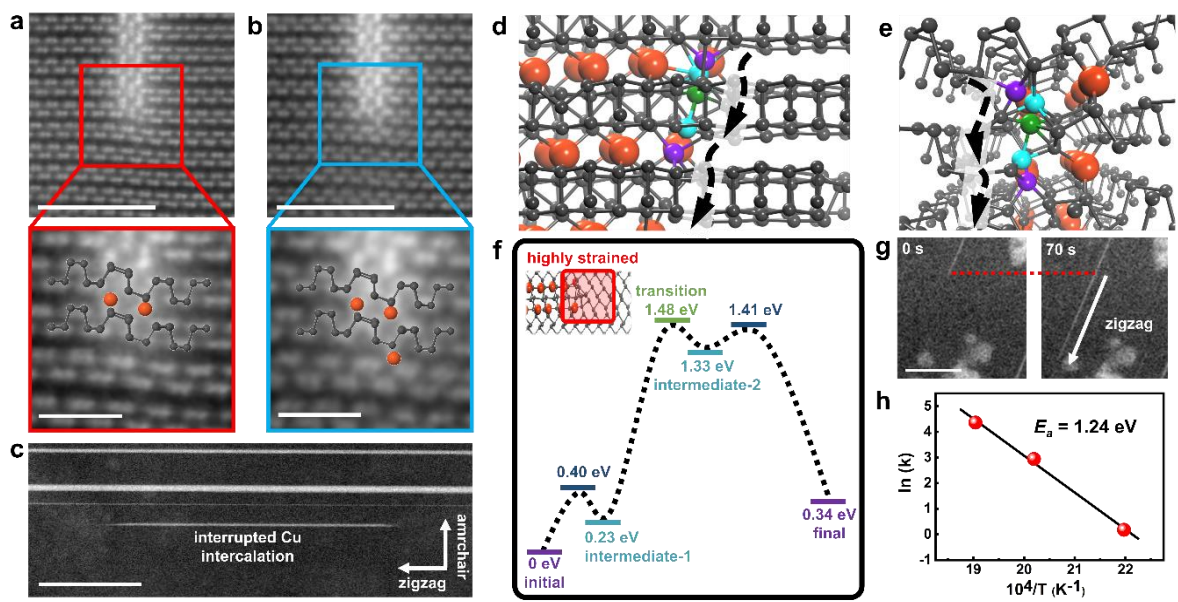

**Fig. 4 | Mechanism of Cu intercalation in BP: Top-down intercalation as the rate limiting process. a-c**, HAADF-STEM images of interrupted Cu intercalated structures in cross-sectional view (**a**, **b**) and plan-view (**c**). Scale bars are 3 nm (upper figures of **a**, **b**), 1 nm (lower figures of **a**, **b**), and 100 nm (**c**). **d-f**, Illustrations (**d**, **e**) and corresponding energy barriers (**f**) of top-down Cu intercalation of strained BP determined from DFT calculations. **g**, HAADF-STEM images of Cu intercalation taken during *in situ* heating at 180 ºC. Scale bar is 50 nm. **h**, Diffusion rate (log-scale) as a function of inverse of temperature obtained from *in situ* heating experiments.

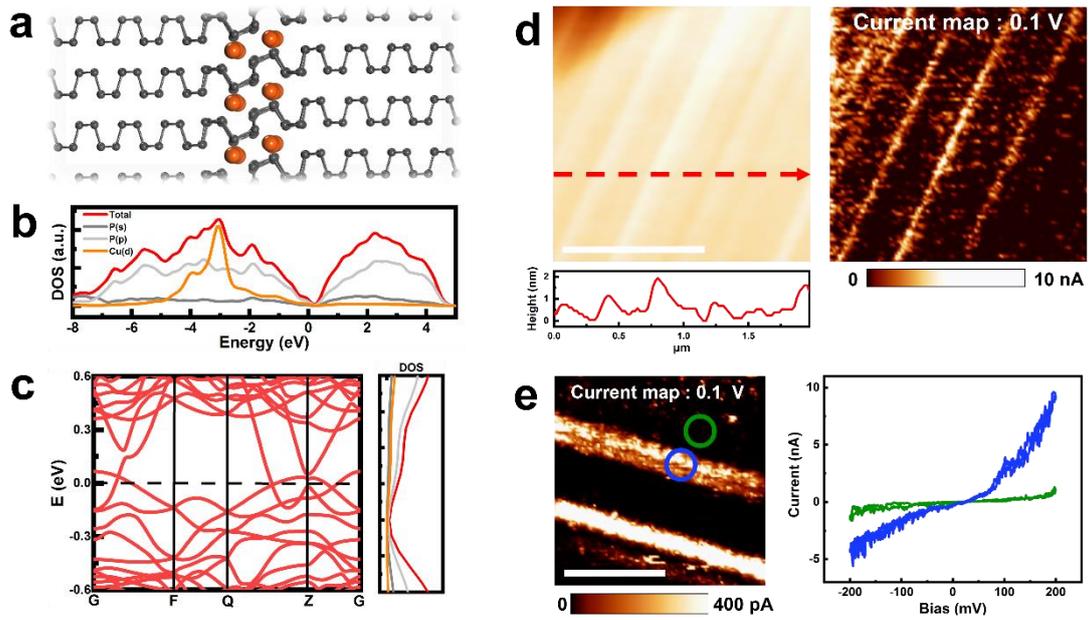

**Fig. 5 | Angstrom-wide conductive channel in Cu intercalated BP. a**, Atomic model used for computational calculation of the density of states (DOS) and the electronic band structure. **b,c** calculated DOS (**b**) and electronic band structure (**c**) with the atomic model. **d**, Topography and current maps of Cu intercalated BP. Scale bar is 1 μm. **e**, a magnified current map (left). Scale bar is 100 nm. *I-V* curves at the location of Cu intercalated region (blue) and BP region (green).